\documentstyle[prl,aps,epsf]{revtex}
\begin{document}
\draft
\twocolumn[\hsize\textwidth\columnwidth\hsize\csname
@twocolumnfalse\endcsname
\title{Commensurate and Incommensurate  
Vortex  States in Superconductors with 
Periodic Pinning Arrays}
\author{C.~Reichhardt, C.~J.~Olson, and Franco Nori}
\address{Department of Physics, The University of Michigan,
Ann Arbor, Michigan 48109-1120}
\date{\today}
\maketitle
\begin{abstract}

As a function of applied field, we find a rich variety of ordered 
and partially-ordered vortex lattice configurations in systems with 
square or triangular arrays of pinning sites. We present formulas that 
predict the matching fields at which commensurate vortex configurations 
occur and  the vortex lattice orientation with respect to the pinning 
lattice. Our results are in excellent agreement with recent imaging 
experiments on square pinning arrays 
{[K.~Harada {\it et al.\/}, Science {\bf 274}, 1167 (1996)]}.
\end{abstract} 
\vskip1pc]
\narrowtext

\section{Introduction}

The concept of an elastic lattice interacting with a rigid substrate 
lattice producing commensurate and incommensurate transitions when 
the periodicities of the two lattices match or mismatch	
is found in numerous condensed matter systems, including atoms adsorbed 
on surfaces \cite{ref1}, layered superconductors \cite{ref2}, 
superconducting networks and  Josephson-junction arrays \cite{ref3}, 
colloids \cite{ref4,ref5}, and magnetic bubble arrays interacting with 
patterned substrates \cite{ref6}. 
Recently, increased interest has been focused on superconducting systems  
with well defined square or triangular periodic pinning arrays  
in which vortices can be trapped both at individual pinning sites and 
also at the interstitial regions between pinning sites 
\cite{ref7,ref8,ref9,ref10,ref11,ref12,ref13}. 
Due to the interstitial pinning, the vortex system 
differs significantly from systems such as atoms on surfaces or 
Josephson-junction arrays. For instance, in the latter systems, 
the potential substrate has an egg-carton form, so the atom or 
vortex lattice has the {\it same\/} configuration whenever it matches 
the underlying potential.  Recent direct imaging experiments \cite{ref7} 
and simulations \cite{ref12} indicate that ordered 
vortex configurations in samples with periodic pinning  can 
{\it vary\/} at each matching field, producing a remarkable variety 
of stabilized vortex lattices which are quite distinct from those 
found in superconducting networks \cite{ref3} and other systems. 

Since highly ordered commensurate lattices can be more strongly 
pinned than incommensurate lattices \cite{ref7,ref9,ref10,ref11,ref12}, 
a determination of how different matching configurations
affect the overall pinning of the vortex lattice could be useful 
for technological applications of superconductors. For instance, 
enhanced pinning at certain matching fields has been verified 
with the observation of peaks in easily measurable magnetization 
curves \cite{ref9,ref11,ref12}, including high-$T_{c}$ 
materials \cite{ref11}. 

Imaging experiments have so far only probed up to the fourth 
matching field and have only examined square pinning arrays 
\cite{ref7}. A general characterization of the vortex matching 
patterns as a function of arbitrary matching densities for square 
and triangular pinning arrays has not been done up to this point. 

We have performed a series of large scale simulated annealing as well as   
flux-gradient-driven \cite{ref14} molecular dynamics simulations of 
vortices interacting with square and triangular arrays of small 
pinning sites for very high fields (up to the 28th matching field), 
and for a wide range of pinning parameters and system sizes. 

Our results show that the vortex lattice (VL) is highly ordered only 
at certain matching fields (MFs) and can have various orientations 
with respect to the underlying pinning array. At some MFs the VL 
is actually disordered.  
The enhancements of $M(H)$ are most noticeable 
for fields less than the second matching field; 
however, we find some 
evidence of small enhancements of $M(H)$ 
for higher fields. 
Square and 
triangular arrays produce {\it different} sequences of ordered matching 
fields at which the pinning is enhanced. At some MFs, we find 
novel vortex arrangements with translational order only along 
certain directions.  Our numerical results are in excellent agreement 
with recent low-field experiments on square pinning arrays \cite{ref7}.
Moreover, using geometrical arguments that take into account the constraints 
of the pinning array, we derive simple formulas for the ordered 
MFs and for the orientation of the VL with respect to the square 
or triangular pinning array. 

\section{Simulation}

We perform simulated annealing simulations for a
2D transverse slice (in the $x$--$y$ plane) of a
superconducting 3D slab   
by numerically integrating the overdamped equation of motion of rigid 
2D vortices: 
\begin{equation}
 \eta {\bf v}_{i} =  
{\bf f}_{i} = {\bf f}_{i}^{vv} + {\bf f}_{i}^{vp} + {\bf f}_{i}^{T}. 
\end{equation}
The term ${\bf f}_{i}$ is the total force 
per unit length acting on vortex $i$. 
The force due to the interactions with other vortices is 
\begin{equation}
{\bf f}_{i}^{vv} = \ \sum_{j=1}^{N_{v}} \ f_{0}\, K_{1}\,\left(
\frac{|{\bf r}_{i} - {\bf r}_{j}|}{\lambda}\right)\,{\bf {\hat r}}_{ij} 
\end{equation}
where $N_{v}$ is the number of vortices,
${\bf {\hat r}}_{ij} = ({\bf r}_{i} - 
{\bf r}_{j})/|{\bf r}_{i} - {\bf r}_{j}|$, and we take $ \eta = 1$. 
$K_{1}(r/\lambda)$ is the modified Bessel function, $\lambda$ is the
penetration depth, and
$$
f_{0} = \frac{\Phi_{0}^{2}}{8\pi^{2}\lambda^{3}} 
$$
The Bessel function decays exponentially for $r$  greater than $\lambda$, 
so for computational efficiency the interaction can be safely cut off
at $ 6\lambda$. In thin-film superconductors the long-range
vortex-vortex interaction decays as $1/r$ unlike in 3D bulk 
superconductors;  however, the excellent agreement between our 
results and  experiments in thin films \cite{ref7} 
indicates that our results are valid for  both slabs and thin 
films and are general enough to be applicable to other 
systems with repulsive particles on a periodic substrate 
(e.g., colloids).
 The pinning force is
\begin{equation} 
 {\bf f}_{i}^{vp} = 
 \ \sum_{k=1}^{N_{p}}\left(\frac{f_{p}}{r_{p}}\right)\, 
|{\bf r}_{i} - {\bf r}_{k}^{(p)}|
\ \Theta 
\left(\frac{r_{p} - |{\bf r}_{i} - {\bf r}_{k}^{(p)}|}{\lambda}\right)
\;{\bf {\hat r}}_{ik}^{(p)}
\end{equation} 
where $\Theta$ is the Heaviside step 
function, $f_{p}$ is the maximum pinning force, $N_{p}$ is the number of 
pinning sites and $ {\bf {\hat r}}_{ik}^{(p)} = 
({\bf r}_{i} - {\bf r}_{k}^{(p)})/|{\bf r}_{i} - {\bf r}_{k}^{(p)}|$.
Temperature is added as a 
stochastic term with properties 
\begin{eqnarray}
< f_{i}^{T}(t)>\, = \,0
\end{eqnarray}   
 and 
\begin{eqnarray}
 <f^{T}_{i}(t)f^{T}_{j}(t^{'})>\, = \, 
2\,\eta k_{B}\,T\, \delta_{ij}\,\delta(t - t^{'}). 
\end{eqnarray}
To find the vortex ground state, 
we gradually cool a fixed number of randomly moving vortices 
from a high temperature to $T = 0$, simulating the field-cooled 
experiments of Ref.\cite{ref7}. 
To examine vortex mobility and 
features in the magnetization curves as a function of applied field 
$H$, we use flux-gradient-driven simulations in which
only the central $2/3$ of the sample contains pinning sites. 
In this case, vortex lines are slowly added to the unpinned region 
and force their way into the pinned region (the actual sample).     
Although the vortex system in  flux-driven simulations is in a 
non-equilibrium state, almost all of the vortex states found by 
simulated annealing also appear in the flux-driven case in parts 
of the sample.

We measure all lengths in units of $ \lambda$ and fields in 
$\Phi_{0}/\lambda^{2}$, and consider systems 
from $36\lambda\times36\lambda$ up to 
$72\lambda\times72\lambda$ in size.
The pinning is placed in square or triangular arrays at densities 
between $n_{p} = 0.072/\lambda^{2}$ and $0.81\lambda^{2}$.  
The pinning radius is fixed at $r_{p} = 0.35\lambda$. 
Pinning sites this size and smaller trap only {\it one} vortex 
per pinning site,
which is similar to the experimental situation in Ref.\cite{ref7}.  
We consider pinning forces  $f_{p}$ varying from $ 0.2f_{0}$ to 
$ f_{0}$ and examine
the VL ordering up to the 28th MF. 

\section{Vortex Lattice ground states: Square Case}   
\subsection{Patterns Experimentally Observed}

In Fig.~1 we show a series of VL orderings after annealing 
from our simulations for a 
square pinning array with $B = 0.17\Phi_{0}/\lambda^{2}$  
for each integer MF up to the 9th MF. In (a), at the first MF, 
all the vortices are trapped at the pinning sites so that the 
overall VL is square. At the second MF (b) the interstitial vortices 
occupy the regions in between the pinning sites, so the overall VL 
is square but rotated $45^{\circ}$ with respect to the pinning array. 
At the 3rd MF (c), the VL is still highly ordered, with pairs of 
interstitial vortices alternating in position. In (d), at the fourth 
MF, a VL with triangular ordering is observed.  These structures 
for the first four MFs correspond  exactly to those found in direct 
imaging experiments \cite{ref7}. We also observe ordered VLs 
at the sub-MFs ($B/B_{\phi} = 1/4$ and $1/2$, 
where $B_{\phi}$ is the vortex density at the first MF), and 
partially ordered VLs at fractional MFs 
($B/B_{\phi} = 3/2$ and $B/B_{\phi} = 5/2$)  
in agreement with experiment \cite{ref7}. 
We find that the 
general features of the observed VL configurations up to the fourth 
matching field are robust for a wide range of parameters 
with $\,0.2f_{0} \leq  f_{p} \leq f_{0}$, and also for 
$0.072\,\Phi_{0}/\lambda^{2} \leq  B_{\phi}\,\leq 0.81\,\Phi_{0}/\lambda^{2}$, 
for      
system sizes up to $72\lambda\times72\lambda$. 

\subsection{Patterns Not Yet Experimentally Observed} 

In Figs.~1(e--i) we show vortex configurations 
from our simulations
that have not yet been 
observed experimentally. In (e), at the fifth MF, the overall VL 
is again square and rotated $27^{\circ}$ with respect to the pinning 
lattice. In (f) a very unusual VL is observed; although the VL is 
neither square nor triangular  some ordering is still visible. 
Along the $(-1,1)$ direction the vortices are spaced periodically 
while in other directions apparently periodic distortions can 
be clearly seen. At the 7th MF, in (g), the VL is disordered. 
In (h), at the 8th MF, the VL is nearly triangular. 
In (i), at the 9th MF, a distorted square VL  with two 
different orientations appears, separated by a twin boundary in 
the middle of the figure. For similar systems with lower pin density 
we have studied up to the 28th MF. We see the same VLs already 
described  as well as ordered VLs at the 12th and 15th MFs, 
while the vortex configurations at the other MFs have no particular 
ordering.  For $B/B_{\phi} > 15$, at high MFs with no overall lattice 
order, the VL contains ordered domains separated by grain boundaries 
of defects similar to those observed in Ref.\cite{ref8}.    

\twocolumn[\hsize\textwidth\columnwidth\hsize\csname
@twocolumnfalse\endcsname
\begin{figure}
\centerline{
\epsfxsize=6.4in
\epsfbox{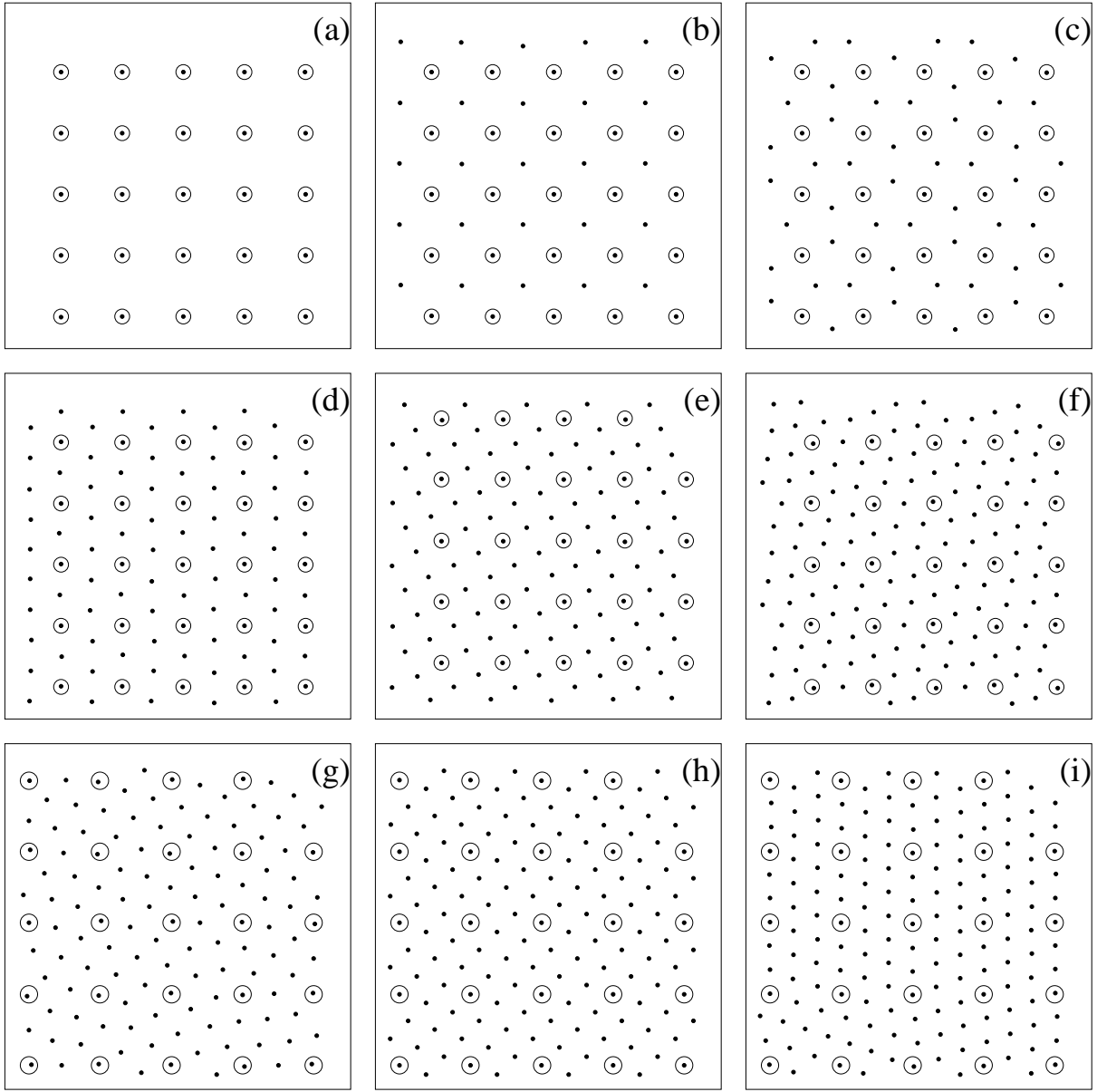}}
\caption{Vortex ground states obtained from simulated annealing for a square 
pinning array with $B = 0.17\,\Phi_{0}/\lambda^{2}$, $r_{p} = 0.35\lambda$, 
and $f_{p}/f_{0} = 0.625$ showing a $12\lambda \times 12\lambda$ subset 
of a $36\lambda \times 36\lambda$ sample. The flux density is 
$B/B_{\phi} = 1$ in (a), $2$ (b), $3$ (c), $4$ (d), $5$ (e), 
$6$ (f), $7$ (g), $8$ (h), and $9$ (i).}
\label{fig1}
\end{figure}  
\vskip2pc]  

Due to numerical constraints we could only look at 
pinning 
densities up 
to $n_{p} = 0.35\Phi_{0}/\lambda^{2}$ for 
$5 < B/B_{\phi} \leq 12$, and 
$n_{p} = 0.072\lambda^{2}$ for matching fields 
$12 < B/B_{\phi} \leq 28$. The vortex 
patterns observed here are robust for system sizes up to 
$72\lambda \times 72\lambda$. The fact that the same patterns appear 
for different-sized systems indicates that 
the patterns arise due to  
commensurability 
with the pinning lattice rather than commensurability with 
the periodic boundary conditions. 
We should point out that since we cannot do infinite-size systems, 
we cannot  conclusively rule out finite size effects on the vortex patterns 
observed. Also, in an experimental sample, edges and line/planar defects 
might distort an otherwise periodic VL and create ordered domains that 
do not extend over the entire sample. 

\twocolumn[\hsize\textwidth\columnwidth\hsize\csname
@twocolumnfalse\endcsname
\begin{figure}
\centerline{
\epsfxsize=6.4in
\epsfbox{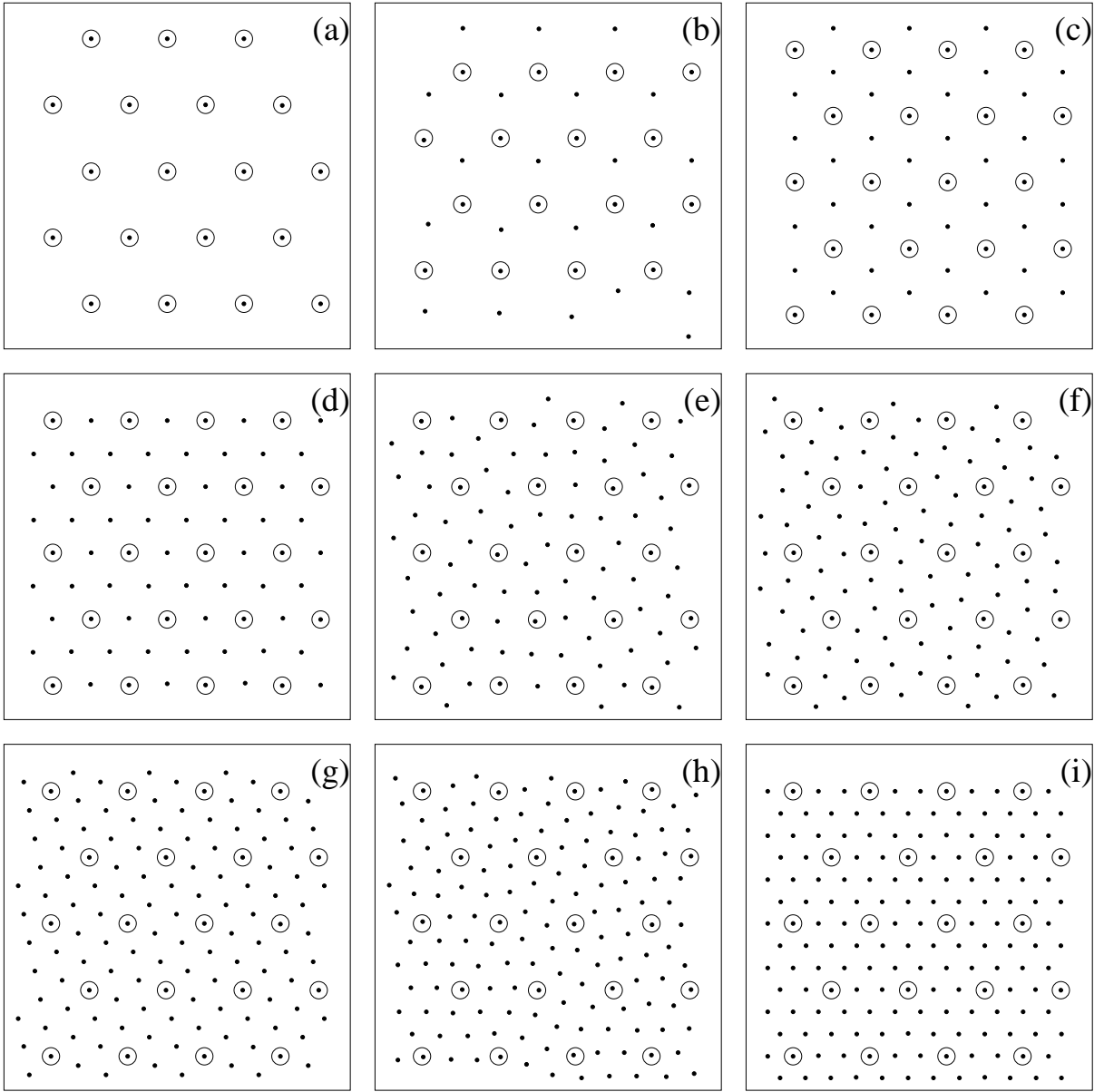}}
\caption{ 
Vortex ground states obtained from simulated annealing for a triangular 
pinning array with  $B = 0.17\,\Phi_{0}/\lambda^{2}$, $r_{p} = 0.35\lambda$,
and  $f_{p}/f_{0} = 0.625f_{0}$ for a $12\lambda \times 12\lambda$ 
subset of a $36\lambda \times 36\lambda$ sample. The flux density 
is $B/B_{\phi} = 1$ in (a), $2$ (b), $3$, (c), $4$, (d), $5$, 
(e), $6$ (f), $7$ (g), $8$ (h), and $9$ (i).
}
\label{fig2}
\end{figure} 
\vskip2pc]

\section{Vortex Lattice Ground States: Triangular case}  

In Fig.~2 we show VL configurations up to the 9th MF for a system 
with the same parameters as in Fig.~1 but with a triangular array 
of pins.  In (a) at the first MF, all the vortices are pinned in a 
triangular lattice.  In (b) a honeycomb VL forms, with some defects 
present. Ordered triangular VLs are seen at the 3rd (c) and 4th 
(d) MFs with the VL rotated $30^{\circ}$ and $0^{\circ}$ respectively 
in relation to the pinning array.  No clear ordering is seen at the 
5th (e), 6th (f), and 8th (h) MFs.  A triangular VL rotated 
$18^{\circ}$ with respect to the pinning array is seen at the 7th (g) 
MF.  In (i) at the 9th MF the VL is triangular and not rotated with 
respect to the pinning array. As in the samples with square pinning, 
we observe ordered VLs at certain sub-MFs but the VL is disordered at 
other non-MFs. The matching vortex configurations are very robust for 
all the parameters we have investigated, except for the honeycomb VL 
which disappears for weak pinning, $f_{p} < 0.3f_{0}$.
For different samples ($n_{p} = 0.072\,\Phi_{0}/\lambda^{2}$,
$f_{p} = 0.625f_{0}$) we have studied up to the 28th MF, and find ordered 
triangular VLs at the MFs of order 12, 13, 16, 19, 21, 25 and 28. 
The vortex patterns observed for the triangular
pinning array are robust for a similar set of parameters as the square 
pinning array discussed in the previous section.

\section{Matching Conditions} 

To derive formulas for the fields at which VLs will be ordered, for
a triangular array of pins, first take a 
matching field $N$ and  
consider any two pinning sites along a symmetry axis that have $N$ 
vortices between them. Now find a third pinning site that forms 
a $60^{\circ}$ angle with the original two pinning sites, 
so all three vertices form an equilateral triangle. 
The distances between the third site and each of the other two are
integer multiples of the pinning lattice constant $a$ where 
\begin{eqnarray} 
a = 1.075\sqrt{\frac{\Phi_{0}}{B_{\phi}}}.
\end{eqnarray}
We label these integers $n$ 
and $m$. The distance between vortices $a_{N}$ at a field $B$ is
\begin{eqnarray}
a_{N} = 1.075\sqrt{\frac{\Phi_{0}}{B}}.
\end{eqnarray} 
By using the law of cosines 
the distances must obey 
\begin{eqnarray} 
(Na_{N})^{2} = (ma)^{2} + (na)^{2} + 2mna^{2}\cos(60^{\circ}). 
\end{eqnarray}
At the MFs, $B = NB_{\phi}$, 
so that $ a_{N} =a/\sqrt{N}$. Substituting this into Eq.~(8) gives 
\begin{equation}
N = m^{2} + n^{2} + nm \; . 
\end{equation}     
This equality predicts that for a triangular array of pins, an ordered
VL will form at values $ N = 1,3,4,7,$ and $9$, exactly as seen in 
our simulations in Figs.~2(a--i).  Equation (9) also predicts the 
higher MFs  ($ N = 12,13,16,19,21,25$ and $28$) that
we have observed numerically.
The honeycomb VL seen in Fig.~2(b) is not predicted
by Eq.~(9) because 
Eq.~(9) only predicts when triangular VLs occur. 
From Eq.~(1) we find that the 
angle the VL makes with respect to the pinning array is 
\begin{equation}
\theta = \arctan \left(\frac{\sqrt{3}m/2}{n + 1/2}\right) \; . 
\end{equation}
This equation indicates that for  
\begin{equation} 
B =n ^{2}B_{\phi},
\end{equation} 
when $m = 0$, the VL is not rotated with respect to the array 
of pins. This is observed for
$N = 1,4,$ and $9$ [shown in Figs.~2(a), 
(d), and (i)],  
as well as for $N = 16$ and $25$. For MFs $ N = 3$,  
where $n=1$, $m=1$, and $N = 7$, where $n=1$, $m = 2$, Eq.~(9) gives 
$\theta = 30^{\circ}$ and $19.11^{\circ}$ respectively, in 
agreement with the VLs shown in Figs.~2(c,g). We have found that 
Eq.~(10) is valid at least  up to the 28th MF studied in our simulations. 

We can derive similar conditions for the square pinning array, 
predicting that ordered VLs appear at the $N$th MF when 
\begin{equation} 
N = m^{2} + n^{2} \;  
\end{equation}
This equation predicts square VLs for $N = 1$, $2$, $4$, $5$, $8$ and $9$. 
Indeed, ordered VLs are seen at these fields (see Fig.~1). 
However, only $N = 1$, $2$, and $5$ are square in the simulation.  
Moreover, the angle of the VL with respect to the array of pins is 
in principle expected to be 
\begin{equation} 
\theta = \arctan{\left(\frac{m}{n}\right)} \;
\end{equation}
These matching conditions 
do not always predict the right VL ordering observed in simulations.  
For instance, the VLs seen at higher fields $N > 9$ have 
triangular or distorted square rather than square ordering.
These equations 
fail when the VL tendency to remain triangular dominates the 
tendency of the pin array to force a square ordering on the VL. 
This is particularly clear for higher fields, $N > 9$, 
when the many interstitial vortices are free to 
minimize their energy by forming triangular lattices.
Equation (9) for the triangular array of pins does not 
have this limitation because {\it both\/} the sample 
and the VL favor a triangular order. 

As we have seen from the simulations, the vortex lattice is 
ordered at the matching fields where the commensurability conditions 
outlined above are met and generally disordered where they are not.
Several low matching fields where these conditions are not 
met still produce  
ordered or partially ordered lattices such as the 
honeycomb lattice at the second matching field for the triangular pinning 
lattice and the alternating interstitial lattice at the 3rd field for the
square array.
This ordering at fields not met by our commensurability conditions 
may occur due to the pinning being more 
dominant at lower fields so that ordering can be imposed on the 
interstitial vortices.
For higher fields the vortex configuration for fields where 
commensurability conditions are not met 
is disordered or partially 
disordered. 
At higher fields $ B > 6B_{\phi}$ the 
vortex-vortex interactions dominate. Here a triangular vortex 
lattice is always preferred, so any alternate ordering 
imposed by the pinning does not occur. 

\twocolumn[\hsize\textwidth\columnwidth\hsize\csname
@twocolumnfalse\endcsname
\begin{figure}
\centerline{
\epsfxsize=6.4in
\epsfbox{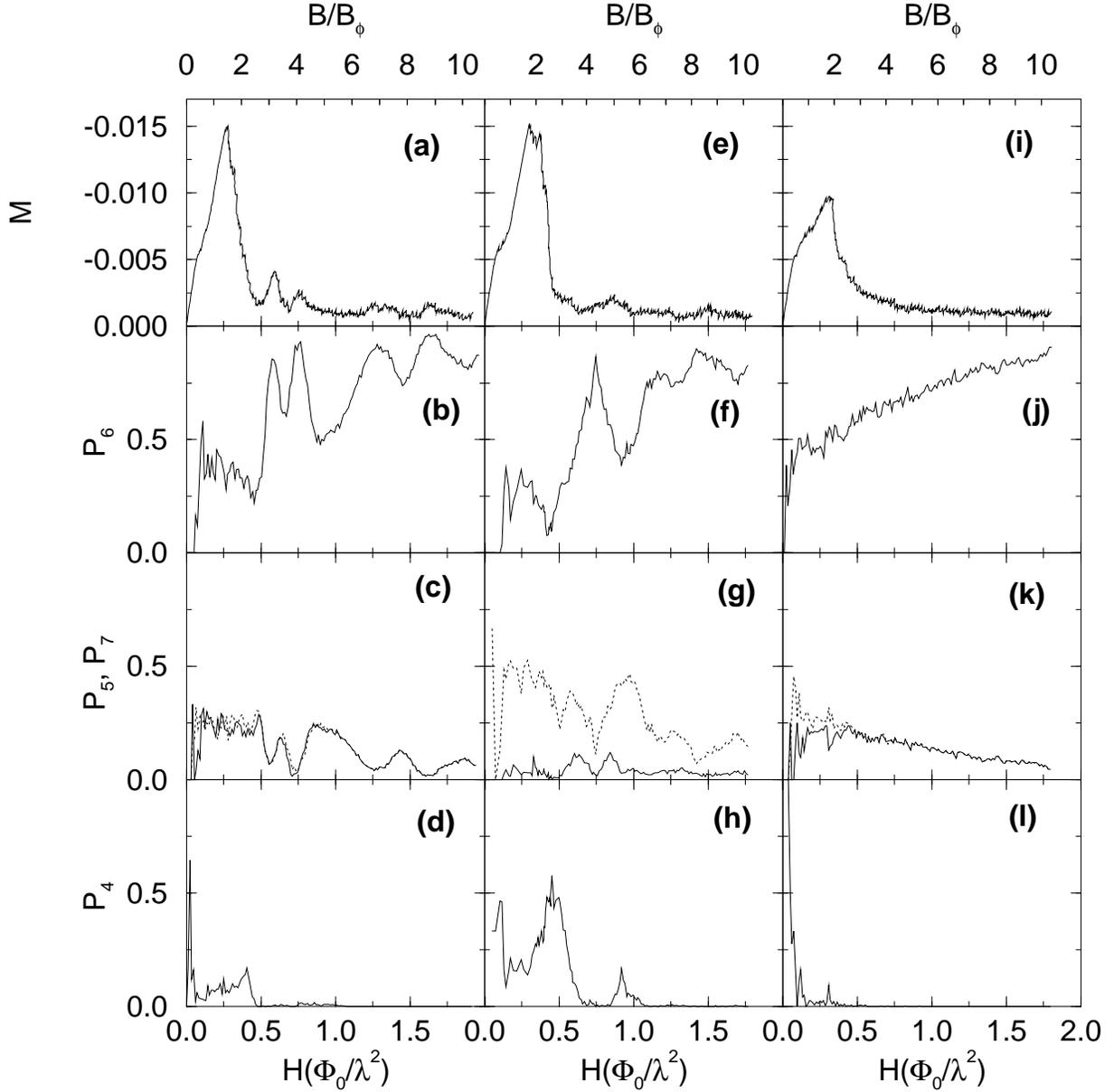}}
\caption{ 
Magnetization $M(H)$ and the fraction of vortices with coordination 
number $k$,  $P_{k}(H)$, for triangular (a--d), square (e--h), and random
(i--l), 
arrays of pins with the same pinning parameters as found in Figs.~1 and 2.  
In (c) and (g), $P_{5}$ ($P_{7}$) is represented with a dotted (solid) line.
In (a) peaks in $M(H)$ can be seen at the 3rd, 4th, 7th and 9th MFs, 
which coincide with peaks in $P_{6}(H)$ shown in (b). For the square 
pinning, small peaks are seen in $M(H)$, (e), at the 2nd, 4th, 
5th and 8th MFs. 
The 4th and 5th (small) peaks merge near the 5th MF. 
The peaks in $P_{6}$, (f), indicate that the VL is 
triangular at the 4th, 6th and 8th MFs, while peaks in $P_{4}$, 
(h), indicate that the VL is square at the 2nd and 5th MFs. 
Sometimes, the peak is slightly shifted 
(e.g. to 2.5 instead of 2 or 4.5 instead of 4) because of
the gradient in the fields. 
In (g) $P_{5}$ is the upper curve, while in (c) and (k) 
$P_{5}$ and $P_{7}$ follow each other.   
For  
the random pinning array at low fields 
the maximum value $|M(H)|$ is $0.0095\,\Phi_{0}/\lambda^{2}$,  
about 1.5 times less than the triangular or square pinning arrays (a,b).   
For $ H > 0.4\,\Phi_{0}/\lambda^{2}$, 
the magnetization $M(H)$ falls off smoothly while 
$P_{6}$ slowly increases as vortex-vortex interactions dominate at 
higher fields. 
}
\label{fig3}
\end{figure}
\vskip2pc]

\section{Topological Order and Magnetization}

In order to relate the MF configurations to vortex mobility as 
well as to  experimentally measurable bulk quantities we 
present the magnetization $M(H)$ 
obtained from flux-gradient-driven simulations \cite{ref12} 
of samples with the 
same pinning parameters used in Fig.~2. In Fig.~3 we plot  
 $P_{k}(H)$, the fraction of vortices with 
coordination number $k$ obtained from the Voronoi construction,
and $M(H)$, a useful and common measurement of the net 
critical current. 
For a perfect triangular lattice $P_{6} = 1$, so any departure from 
this indicates a defective lattice.   A peak in $M(H)$ indicates 
enhanced pinning. In Fig.~3(a) we find that $M(H)$ is very large 
for fields less than about $2B_{\phi}$ and falls off very rapidly 
after this.  Peaks in both $M(H)$ and $P_{6}(H)$ appear
[Figs.~3(a,b)] at the MFs $ N = 3, 4, 7,$ and $9$ that produced 
triangular VLs in the simulated annealing [Fig.~(2)]. 
From Fig.~2(a),  
the vortex lattice would be expected to form 
a triangular lattice with the pinning 
substrate at the first matching field, 
and $P_{6}$ would be expected
to be equal to one. 
In the flux-gradient-driven case shown in Fig.~3(a,b),
no peak in $M(H)$ or $P_{6}$ near the first matching field
is observed.
This is due to the fact that the 
large {\it flux gradient} at low 
fields strongly distorts the VL. In the field cooled situation shown in 
Fig.~2(a) 
there is no gradient in the vortex density to 
interfere with the vortex lattice ordering.
For the flux-gradient-driven case at  
higher fields, $B > 2B_{\phi}$, 
the gradient flattens so that the vortex 
lattice can become commensurate with the pinning substrate over a large area. 
We note that even for high matching fields  
a small flux-gradient will always be present so that 
$P_{6}$ will always be less than one as seen in Fig.~3(b).
For weaker pinning,  
$f_{p} \leq 0.3f_{0}$, the flux gradient is reduced at
low fields so that a peak in $M(H)$ and $P_{6}$ can be observed at 
$B/B_{\phi} = 1$ 
\cite{ref12}. We find that this behavior is independent of system size. 
At the MFs $N = 5$ and $N = 6$ the VL is highly 
defective with $P_{6}(H)$ dropping as low as $0.5$.   
No peak appears in $ M(H)$ 
for the 2nd MF. For systems in which we have studied up to the 
28th MF we also see some enhancements in $M(H)$ and 
$P_{6}(H)$ at the MFs predicted by Eq.~(9), 
although the features are washed out at high fields.    

In Fig.~3(e) we show $M(H)$ for a square pinning array with the same 
parameters used in Fig.~1. Again $M(H)$ is large for low fields       
and rapidly  falls off after the 2nd MF. We can see 
a dip after the 3rd MF and an overall enhancement in 
$M(H)$ at the 2nd, 4th,  5th, and 8th MFs, although 
no clear enhancement is seen at the 6th and 9th MFs even 
though the VLs observed through simulated annealing
at these fields  also appear in this flux-
gradient driven simulation. 
To examine the evolution of the vortices with four nearest-neighbors 
we consider a slightly modified Voronoi algorithm in which
the lengths of each side of a Voronoi cell are compared. If the 
length of any side is less than one-fourth of the average lengths of the 
other sides, then it is ignored.  
$P_{4}(H)$ first shows a peak at the second MF 
when the vortices form the lattice shown in Fig.~1(b).
There is no peak in $P_{4}(H)$ at the first MF due to
the large flux gradient. 
$P_{6}(H)$ shows a large peak at the 4th MF that 
corresponds to the triangular VL seen in Fig.~1(d).
$P_{6}(H)$ then 
drops rapidly and $ P_{4}(H)$ increases as the VL gains the 
square ordering seen in Fig.~1(e). $P_{6}(H)$ rises 
at the 6th MF and peaks at the 8th.   
In square pinning arrays, where we have gone up to the 28th MF, 
small enhancements of $M(H)$ are observed for most of the  
MFs that produced ordered VLs. 
The results indicate that,
without directly imaging the VL, it could be experimentally 
possible to deduce the existence of the ordered vortex arrays seen 
here, by looking for a specific sequence of peaks in $M(H)$, at 
least up to the 5th matching field.
Beyond the fifth matching field we observe only very small peaks 
in $M(H)$, which may make them difficult to see experimentally. 

Our results
are only valid for pins small enough that 
{\it only one\/} vortex can be trapped in each pinning site.  
With triangular pinning,  
peaks in $M(H)$ should
in principle
occur for MFs 
$N$ that satisfy Eq.~(9). For square pinning arrays, we observe that 
peaks in $M(H)$ 
occur for MFs given by 
$ N = n^2 + m^2 $, when $ N \leq 10$, 
and by 
$ N = n^2 + m^2 - 1$, when $ N > 10$.
This pattern of peaks differs from those already seen
experimentally using periodic pinning arrays with 
large pinning radii, as first shown in Ref.\cite{ref12}. In 
experiments, peaks in $M(H)$ are usually observed at every MF  
due to {\it multiple\/} vortices being trapped in pinning sites 
\cite{ref9,ref11}. 

To compare the effects of random pinning to square and triangular arrays, 
in Fig.~3(i) 
we plot $M(H)$, and in Fig.~3(j-l) we plot 
$P_{k}(H)$ for a sample with the same 
pinning parameters as in Fig.~3(a,e) 
except the pinning sites are placed in 
a random array. It can clearly be seen that most of the peaks in 
$P_{k}$ and $M(H)$ are washed away with $ M(H)$ having a smooth falloff
after the peak and the strong variations in $P_{k}$ lost. The 
fraction of $P_{6}$ gradually increases as the vortex-vortex interactions 
dominate at higher fields. 
At the matching fields, the random array of pins has no peaks or 
enhancements in $M(H)$ or $P_{k}$. This suggests that the presence of 
peaks in the periodic pinning arrays are due to the commensurability effects 
with the pinning substrate.

The maximum value of 
the absolute value of  
$M(H)$, $|M(H)|$,  
is 
$ 0.0095\,\Phi_{0}/\lambda^{2}$ 
for the random pinning in Fig.~3, 
while it is $ 0.015\,\Phi_{0}/\lambda^{2}$ for   
the triangular array and square array. The latter value  
is about 1.5 times larger than that found for the random pinning case.
This enhancement of $M(H)$ occurs 
only for a limited range of fields. The
$M(H)$ for the  
triangular pinning array falls to the 
same value as $M(H)$ for   
the random pinning array 
at $H \approx 0.30\,\Phi_{0}/\lambda^{2}$, which is less 
than $2B_{\phi}$. The $M(H)$ for the square pinning array 
remains higher than the $M(H)$ of the random pinning array 
until $H \approx 0.45\,\Phi_{0}/\lambda^{2}$. 
This higher value of $H$ at which the drop occurs for the square
pinning array
is due to the 
strong commensurability at the 2nd MF for the square
pinning array,
whereas for the triangular array 
the second MF is 
a less stable defective honeycomb 
lattice. For the triangular pinning array 
at the third MF, 
$M(3B_{\phi}) \approx 0.004\,\Phi_{0}/\lambda^{2}$,
 while the random pinning gives 
$M(3B_{\phi}) \approx 0.0025\,\Phi_{0}/\lambda^{2}$. For fields higher than 
the fourth MF, 
$M(H)$ is of the same order for the three 
pinning array geometries studied here.  

\section{Remarks on Finite-size Effects}
Regarding finite size effects, we would like to
emphasize that
we have conducted simulations
in samples that vary in size
from
$36\lambda \times 36\lambda$
up to $72\lambda\times 72\lambda$, and we observe the
same features
in all our simulations regardless of the system size.
We have also done simulations with different
pinning strengths and observe the same peaks in $M(H)$
and $P_{6}(H)$. This
reproducibility in the peaks in different simulations suggest that
the peaks are not merely
fluctuations but are robust and reproducible results.
To further address this issue
we have included in Fig.~3 both $M(H)$ and $P_{k}(H)$
for a system with the same pinning parameters
as in the first two plots of $M(H)$ but with pinning placed
randomly. In this plot no peaks are visible in $M(H)$ beyond the
initial peak nor are any peaks visible in
$P_{k}(H)$.
The same behavior 
for the random array is observed for different sized
systems.
If the peaks in $M(H)$ in systems with square and triangular pinning are
due to finite size effects such as commensurability with the boundary
conditions, then peaks in $M(H)$ and $P_{k}$
for system with the same size and boundary conditions
but with random pinning should appear as well.

The absence of any peaks in
$M(H)$ and $P_{k}$ for the system with random pinning
strongly suggests
that peaks in these quantities for the square and triangular
pinning array are due to commensurability effects with the pinning
lattice only.
It is
important to stress that in our simulations, our analytical results
and
experimentally observed vortex lattice (VL) configurations are all
consistent with each other. 

\section{Conclusion}

To summarize, we have studied VLs interacting with periodic pinning 
arrays in which interstitial pinning is relevant above the first MF.  
We have shown that this system behaves considerably differently from  
atoms on surfaces or Josephson-junction arrays. A rich variety of 
distinct VLs can be stabilized including several novel partially-ordered 
lattices. We have also derived commensurability conditions for MFs 
at which stable ordered VLs appear.  For the triangular pinning array these 
commensurability conditions are in excellent agreement with our 
simulations, while for the square array the commensurability conditions 
work for low fields (up to the 10th matching field).      
Our simulations are in excellent agreement with recent imaging experiments 
\cite{ref7} and are robust over a wide range of parameters and system sizes. 
Our predictions can be tested with Lorentz microscopy techniques, 
Bitter decoration techniques, and by looking for a specific sequence
of peaks in magnetization measurements. These phases should be 
accessible to other systems with periodic pinning, including 
charged colloidal particles in a periodic array of optical traps 
\cite{ref5}, 
and magnetic bubble arrays interacting with 
patterned substrates. 

\section{Acknowledgments}

Computer services were provided by the Maui High Performance Computing
Center, sponsored in part by the Phillips Laboratory, Air Force Materiel
Command, USAF, under cooperative agreement No F29601-93-2-0001. 
Computing services were also provided by the University of Michigan 
Center for Parallel Computing, partially funded by NSF Grant No 
CD-92-14296. C.O. acknowledges support from the NASA Graduate Student 
Researchers Program.

\end{document}